# PACMAN Attack: A Mobility-Powered Attack in Private 5G-Enabled Industrial Automation System

Md Rashedur Rahman*, Moinul Hossain*, and Jiang Xie**
*George Mason University, VA, USA
**The University of North Carolina at Charlotte, NC, USA
Email: mrahma47@gmu.edu, mhossa5@gmu.edu, and jxie1@uncc.edu

*Abstract*—3GPP has introduced Private 5G to support the next-generation industrial automation system (IAS) due to the versatility and flexibility of 5G architecture. Besides the 3.5GHz CBRS band, unlicensed spectrum bands, like 5GHz, are considered as an additional medium because of their free and abundant nature. However, while utilizing the unlicensed band, industrial equipment must coexist with incumbents, e.g., Wi-Fi, which could introduce new security threats and resuscitate old ones. In this paper, we propose a novel attack strategy conducted by a mobility-enabled malicious Wi-Fi access point (mmAP), namely *PACMAN* attack, to exploit vulnerabilities introduced by heterogeneous coexistence. A mmAP is capable of moving around the physical surface to identify mission-critical devices, hopping through the frequency domain to detect the victim's operating channel, and launching traditional MAC layer-based attacks. The multi-dimensional mobility of the attacker makes it impervious to state-of-the-art detection techniques that assume static adversaries. In addition, we propose a novel Markov Decision Process (MDP) based framework to intelligently design an attacker's multi-dimensional mobility in space and frequency. Mathematical analysis and extensive simulation results exhibit the adverse effect of the proposed mobility-powered attack.

## I. INTRODUCTION

The future industrial automation systems (IAS) is envisioned to adopt the full-scale wireless connectivity offered by the fifth-generation (5G) cellular technology [1]. Regulatory authorities and researchers are advocating the implementation of Private 5G in IAS to accomplish the precise industry-specific QoS standards [2], [3]. Due to the limited availability of licensed radio spectrum and the additional logistics required to access such resources, *unlicensed spectrum bands*, such as 5 GHz, have the potential to play a significant role [4]. Though industries may prioritize using licensed spectrum bands as anchor carriers, unlicensed spectrum bands provide an unparalleled resource to meet the demands of advanced AI/ML-enabled IAS applications. The caveat, however, is that Private 5G is required to coexist with the incumbents in these unlicensed bands, e.g., Radar and Wi-Fi in the 5GHz band.

**Motivation:** While Private 5G-enabled IAS plan to utilize unlicensed spectrum bands like 5GHz, the heterogeneity between Wi-Fi and cellular technologies may hinder their fair and effective coexistence. For example, Wi-Fi employs a preamble-based detection mechanism for Wi-Fi signals and an energy-sensing-based detection mechanism for non-WiFi ones, whereas LTE/5G employs the latter. Cellular technologies have adopted CSMA/CA-based methods along with a similar contention window structure and backoff techniques to maintain a uniform channel access framework in 5GHz spectrum band. The CSMA/CA mechanism, however, has several vulnerabilities that may have an adverse impact on IAS applications. Additionally, the absence of a preamble-based detection mechanism makes it more difficult to detect such malicious behaviors. Therefore, given the national security significance of manufacturing and supply chain industries, it is crucial to assess such vulnerabilities and propose a more secure coexistence framework in unlicensed spectrum bands.

**Challenges:** Most research on the fair coexistence of Wi-Fi and cellular technologies in the 5GHz spectrum band has prioritized Wi-Fi's performance. In contrast, only a small amount of research has addressed the Quality-of-Service (QoS) of cellular technologies. Moreover, the impact of various PHY/MAC layer-based attacks in *heterogeneous* spectrum coexistence scenarios is rarely studied. In [5], authors have conducted a comprehensive survey on the vulnerabilities in the future heterogeneous coexistence of 802.11 and cellular technologies in an unlicensed spectrum band. Researchers in [6] and [7] have proposed intelligent jamming attack and MAC layer-based misbehavior, respectively, in the context of spectrum coexistence of Wi-Fi and LTE in the 5GHz spectrum band. However, only [6] has considered cellular technologies as a victim against malicious AP. Evidently, it is an effective approach to use malicious APs to disrupt cellular communication in the 5 GHz spectrum band. Nonetheless, the attack strategies mentioned previously would not be able to make a consequential impact due to the consideration of the fixed physical location of the attacker and not considering delay-sensitive application scenarios—important security considerations for industrial applications. Additionally, a comprehensive attack strategy must account for the risk of exposure and the reward of persistent attacks. Hence, the consideration of latency-sensitive critical application scenarios (e.g., IAS) with added mobility considerations at the attacker (in both space and frequency), introduces novel challenges to design intelligent adversarial strategies and assess associated vulnerabilities.

**Contribution:** Based on the above discussion, in this paper, (i) we propose a Wi-Fi-based mobility-powered attack model called *PACMAN attack*, where the attacker can traverse the physical area in a private 5G-enabled IAS, locate critical

This work was supported by the US National Science Foundation (NSF) under Grant No. 2304668.

areas for IAS operations, and perpetrate MAC-layer attacks on the devices residing in these areas. To the best of our knowledge, this is the first work to propose such an attack in private 5G-enabled IAS; (ii) in addition, we propose a Markov decision process (MDP) based framework for path planning, attack strategy design, and detection avoidance where the attacker trade-offs between the path that maximizes the attack performance and evasive maneuvers that minimizes the risk of detection. The proposed framework aims to aid in modeling and assessing security vulnerabilities.

## II. RELATED WORK

In the following, we discuss the prior research on the path planning model based on the MDP and PHY/MAC layer vulnerabilities of heterogeneous spectrum coexistence.

### A. Path Planning Model

Path planning models are used in applications such as UAVs [8], drones [9], autonomous vehicles [10], and others where an agent must move through an environment while taking actions that are linked to rewards and penalties to accomplish specific goals. MDP has been a popular method for developing these models. In [9], the authors used a combination of Jump Point Search and MDP to propose a 3D path planning model and real-time collision resolution for multi-rotor drones operating in hazardous urban low-altitude airspace. The trajectory planning for UAV-mounted mobile edge computing systems is formulated using the combination of MDP and Reinforcement Learning in [11]. In [8], [10], authors considered the partially observable MDP-based path planning models for military-based UAVs and the detection of hidden road users by autonomous vehicles. However, these models consider the traversal in the physical domain, whereas our proposed model considers the mobility in both the physical and frequency domains. Additionally, in our proposed model, the adversary traverses the physical space to locate the critical devices and hops through channels to detect the victim's operating channel, all while avoiding exposure.

### B. MAC and PHY Layer-based Vulnerabilities

MAC and PHY layer-based vulnerabilities like jamming, selfish backoff attack etc. have been prevalent in wireless communications since its inception. Spectrum coexistence of heterogeneous technologies can bring a new perspective in terms of vulnerabilities and the detection and defense strategies against them. In [6], the author proposed a jamming attack perpetrated by a malicious Wi-Fi AP to degrade the performance of coexisting LTE users. Conventional jamming attack would not be an energy-efficient strategy for a malicious entity with energy constraints [12]. Although a reactive jamming attack is effective and energy-efficient, it suffers from hardware constraints [12]. Moreover, MAC-layer-based security attacks have the potential to disrupt the harmonious coexistence of heterogeneous technologies in unlicensed spectrum bands. Authors in [13] introduced a logistic classification approach to detect Selfish Backoff attacks in IEEE 802.15.4 networks, whereas [14] utilized a supervised learning model for detecting the backoff manipulation attack in cognitive radio. [15] considered time series analysis to detect malicious nodes using the greedy MAC Protocol. However, while detecting malicious actors or behaviors in the network, none of these studies considered the heterogeneity of technologies in a specific spectrum band. Although researchers in [7] have proposed such MAC layer misbehavior for the first time in such scenarios, their work did not address the scenarios of malicious Wi-Fi APs. At the same time, none of the proposed work on this topic considered the attacker's ability to move across the physical surface. Though Wi-Fi AP-based spoofing attacks [16]–[18] also have the potential to impact the coexistence of heterogeneous technologies, the difference in the proposed work lies in the fact that the malicious entity would act as a legitimate user of the spectrum band and have the required mobility throughout the attack surface.

## III. PROPOSED ATTACK STRATEGY

In the PACMAN attack scenario, the attacker has two ways of traversal throughout the attack surface i.e. spatial and frequency. The attacker will divide the physical space into multiple polygons i.e. zones depending on its interference range (we consider each zone a hexagon) and will initially traverse throughout the surface randomly to have a better understanding of the surface. While moving through different zones, the attacker learns about transmitting devices in different zones. We assume that the attacker has an out-of-bound link (i.e., a secure control channel for the attacker only), through which it derives the reward of attack in each zone. The short-term goal is to *cause successive transmission failures* to reach the maximum limit of transmission attempts (or until the information becomes stale) and seize the victims' operation. The long-term goal is to *locate the sectors* crucial to industrial automation, seize its operation, and remain undetected.

### A. MAC Layer Misbehavior Strategy

In the context of this paper, we only focus on the 5GHz unlicensed spectrum band where Wi-Fi acts as an incumbent user and Listen Before Talk (LBT) based access mechanism is promoted by the regulatory bodies for the cellular technologies. In an LBT-based mechanism, cellular technologies are required to adopt a CSMA/CA-based access mechanism while employing an energy detection-based sensing method. Although in different research, the authors used energy detection levels ranging from -62 dBm to -82 dBm, we would only focus on the energy detection level of -72 dBm based on 3GPP specification [19]. To limit the scope, we only focus on the selfish backoff attack approach [14], in which a malicious Wi-Fi AP tries to employ a lower backoff value to gain more access to the channel while restricting other users. The attacker's choice of backoff value is configurable and depends on the attack objective. In a PACMAN attack, the goal is to restrict the victim from accessing the channel and increase the victim's channel access delay, which will affect the operation of IAS. Although selfish backoff attacks have been studied before, the absence of preamble-based mechanisms to detect malicious behaviors and the attacker's mobility (in space and frequency) create an opportunity for adversaries to stay undetected by traditional intrusion detection systems (IDS).

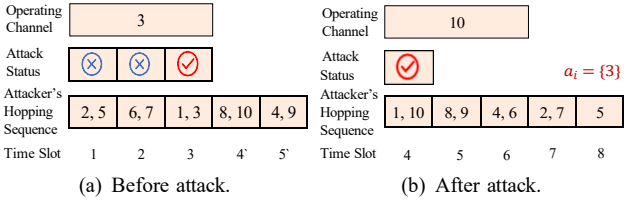

Fig. 1: First phase of the attack.

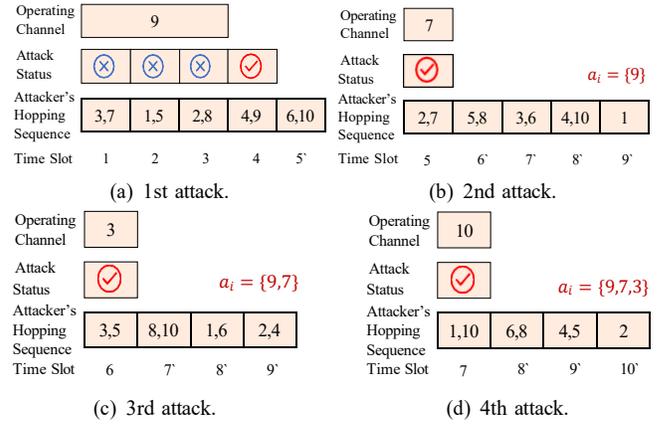

Fig. 2: An illustration of successful DoS attack with $G = 4$.

### B. Frequency Hopping Mechanism

**Short-term Strategy:** A zone has $M$ channels available for the victims device, and the attacker is unaware of the channel that is currently in use. Hence, the attacker visits $m$ different channels during each slot to identify the operating channel. The attacker randomly creates a channel-hopping sequence and periodically hops through it until it locates the victim's active channel. The strategy of channel hopping helps attackers to put an upper bound on how long a victim device can continuously use a channel. Fig. 1(a) shows an illustration of the attack sequence with $M = 10$ and $m = 2$. It shows the hopping sequence of attackers before a successful attack. Here, the operating channel of the victim network in a sector is channel-3 and, in slot-3, the attacker perpetrates the attack. $a_j$ represents the channels where attackers have conducted the attack and $j$ represents the number of successive attacks. After realizing performance degradation, the victim will randomly hop to a new channel, will try to stay on that channel as long as plausible, and will not hop back to the previously attacked channels (i.e., $a_j$) until it achieves a successful transmission. Hence, the attacker will discard the previously attacked channels for a particular transmission attempt. After each successful attack, the attacker randomizes its hopping sequence, excluding $a_j$. Therefore, after $j$ successive transmission failures, attackers have $M - j$ channels to randomize. Fig. 1(b) illustrates a new hopping sequence of attackers where the attacker detects and perpetrates the attack in the first slot.

**Long-term Strategy:** Given the flexibility in the frequency domain, the attacker aims to gain more success in detecting the correct operating channel of the victim. To successfully detect the target victim's operating channel, the attacker has two assumptions. Firstly, when a victim is being denied access to the channel for a continuous period, after a certain threshold it would move to a new channel. Secondly, the victim would not go to a channel where it previously faced anomaly in terms of accessing the channel. When the attacker assuming after $G$ consecutive transmission failures ($G<M$) the victim cancels the current transmission, the attacker stays persistent to increase its chance of successful attack after each successive attacks; hence, it discards earlier attacked channels. Fig. 2 shows an illustration of a scenario, where $G = 4$, and attackers are successful to drop the packet with successive attacks.

After $j^{th}$ successful attack, if the attacker is not successful in the $(j + 1)^{th}$ slot, it assumes that the victim had a successful transmission. Hence, it will re-randomize the hopping sequence (i.e., nullify $a_j$), but exclude the channels it has visited in the current slot (since currently visited channels are free, visiting again is not required), and begin a new period (one period = $\lceil M/m \rceil$ slots).

### C. Physical Mobility

The malicious entity performing a PACMAN attack has the capability of traversing throughout the physical domain of the physical surface. The mobility across the attack surface makes it more difficult for the IDS to locate and identify the selfish or hostile transmissions. To begin, the attack is launched initially in a certain area of the surface, which has a minimal effect on the aggregate but a substantial effect on the zone. Second, while the IDS can detect an abnormal event across the entire surface, differentiating an attack from any physical anomaly in that particular zone can be difficult. Finally, tracking down the attacker's current location as well as its intended future location can be challenging because it may continuously move around the physical surface.

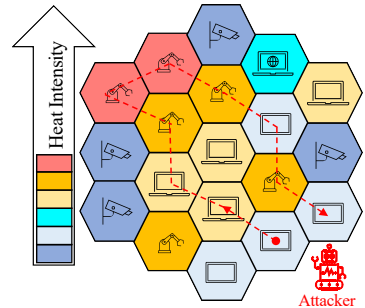

Fig. 3: The physical surface.

Fig. 3 illustrates a physical surface that contains multiple distinct locations in hexagonal shapes comprised of different application areas of the IAS. Also, an estimated travel path of the attacker throughout the surface is shown. The attacker's goal is to identify the operating channel of the victim in a certain location and conduct its MAC layer-based attack to disrupt the network while minimizing the mobility cost. However, the increased duration of attack in a zone would increase the probability of being detected by the IDS. Hence, the attacker moves to an optimal adjacent zone to protect itself from being detected. An optimal location is considered a zone where the attacker would be able to cause the most damage by affecting critical communications. The process of choosing the ideal zone depends on the relative importance of each sector and is discussed in detail in the subsequent section.

**Summary:** The proposed strategy introduces uncertainties in the actions and location of the attacker. Unlike deterministic approaches, the proposed attack strategy introduces a random hopping sequence and path trajectory. Also, the attacker can only be detected if the victim experiences transmission failures; hence, the first attack will always be undetected.

## IV. PROPOSED MDP-BASED ATTACK MODEL

### A. Formation of the MDP

We have utilized the MDP framework proposed in [20], [21] and scaled it to incorporate the mobility of the attacker. We presume that the operating channel of the victim in a location is unknown to the attacker; the attacker iteratively sweeps through the available channels, detects it, and perpetrates a MAC-based attack. As we consider the attacker can sense multiple channels at once (i.e., $m$), instead of waiting on a certain frequency, it will hop through different channels. The attacker will decide its action at the end of each time slot, based on the observation of the current state. The attacker receives an immediate reward $U(n)$ in the $n_{th}$ time slot,

$$U(n) = L.\mathbf{1}(Single\ attack) + Q.\mathbf{1}(Packet\ drop)$$
$$- B.\mathbf{1}(Busy\ channel) - V.\mathbf{1}(Moving\ cost) \quad (1)$$
$$- C.\mathbf{1}(Hopping\ cost) - E.\mathbf{1}(Attacker\ detection)$$

where $\mathbf{1}(\cdot)$ is an indicator function of the event in brackets.

As the employed strategy impacts the current state and also the future states, the expected reward of this game is,

$$\overline{U} = \sum_{n=1}^{\infty} \delta^{n-1} U(n), \quad (2)$$

where $\delta$ represents the discount factor ($0 < \delta \leq 1$). It measures the significance of the future reward values.

### B. Markov Model

This subsection demonstrates the proposed MDP model and defines state space, action space, rewards, and transition probabilities. We assume that the attacker sweep through all channels periodically; hence, the probability of an operating channel being detected depends on the channels that have been visited earlier in the sequence, conforming the requirement of a Markov process (i.e., a future state of the Markov process depends only on the current state).

**Markov States:** The state denotes the status of an attacker at the end of a time-slot at location $l$. Here, the proposed Markov model (Fig. 4) has four kinds of states in each location:
$P^l$ : The attacker senses that the channel is occupied by a PU.
$H_i^l$ : The attacker hopped onto a new channel and had $i$ consecutive unsuccessful detection of the victim ($1 \leq i \leq K$).
$A_j^l$ : The attacker successfully perpetrated $j$ consecutive attacks ($1 \leq j \leq G$).
$D^l$ : The attacker is detected by the IDS system at the site.
We represent the whole state space as $\mathsf{X} \triangleq \{P^l, H_1^l, H_2^l, \cdots, A_1^l, A_2^l, \cdots, D^l\}$, where $l \in \{1, \cdots, L\}$.

For example, $L = 7$ for 7 sector model. In Fig. 4, blacked dotted arrows represent the incoming and outgoing transitions to neighboring locations.

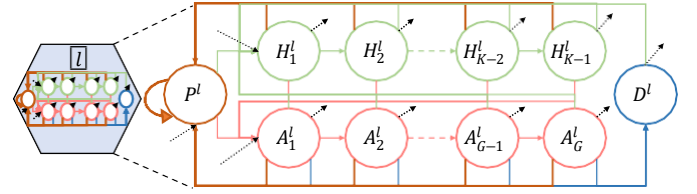

Fig. 4: The proposed MDP-based attack model.

**Actions:** We have three action types available at each state:
$stay_{loc} + hop_{freq}$ ($sh$): The attacker stays at the current location in the next time-slot and hops to the next channels in the hopping sequence.
$move_{loc} + hop_{freq}$ ($mh$): The attacker moves to a new location and hops on new channels.
$move_{loc} + stay_{freq}$ ($ms$): The attacker moves to a new location and stays on the current channel.
We represent the whole action space as $\mathsf{A} \triangleq \{sh, mh, ms\}$.

**Rewards:** Let $U(S, a, S')$ represent the reward when an attacker takes action $a \in \mathsf{A}$ in state $S \in \mathsf{X}$ and enters into state $S' \in \mathsf{X}$. Now using (1), we define rewards:

$$U(S,a,S') = \begin{cases} -C, & \text{if } \{S,a,S'\} = \{\mathsf{X}, sh, H_i\}, i = 1, \cdots, K-1 \\ L-C, & \text{if } \{S,a,S'\} = \{\mathsf{X}, sh, A_j\}, j = 1, \cdots, G-1 \\ Q-C, & \text{if } \{S,a,S'\} = \{A_{G-1}, sh, A_G\} \\ -B-C, & \text{if } \{S,a,S'\} = \{\mathsf{X}, sh, P\} \\ -E-C, & \text{if } \{S,a,S'\} = \{A_j, sh, D\}, j = 2, \cdots, G \\ -V, & \text{if } \{S,a,S'\} = \{\mathsf{X}, ms, H_1\} \\ L-V, & \text{if } \{S,a,S'\} = \{\mathsf{X}, ms, A_1\} \\ -B-V, & \text{if } \{S,a,S'\} = \{\mathsf{X}, ms, P\} \\ -C-V, & \text{if } \{S,a,S'\} = \{\mathsf{X}, mh, H_1\} \\ L-C-V, & \text{if } \{S,a,S'\} = \{\mathsf{X}, mh, A_1\} \\ -B-C-V, & \text{if } \{S,a,S'\} = \{\mathsf{X}, mh, P\} \end{cases}$$
(3)

**Transition Probabilities:** As the attacker can sense $m$ channels at once and go through its attack channel sequence, at state $H_i$, only $\max(M - im, 0)$ channels have yet to be visited by the attacker, and another $m$ channels will be visited in the subsequent slot. Therefore, the probability of an attack (with action $sh$) in the absence of a victim on the channel,

$$Pr_{at|sh} = \begin{cases} \dfrac{m}{M-im}, & \text{if } i < K \\ 1, & \text{otherwise.} \end{cases} \quad (4)$$

We assume a 5G transmission is $q$ mini-slots long. Also, we can approximate the probability of finding the channel busy with action $hop_{freq}$ as the steady-state probability,

$$Pr_{P|a,s} = \dfrac{\alpha}{\alpha+\beta} = \rho, \quad a \in \mathsf{A} \text{ and } s \in \mathsf{X}, \quad (5)$$

where $\alpha$ and $\beta$ represent radar activity and denote transition probabilities from OFF to ON and ON to OFF, respectively. Now, the transition probabilities from state $H_i$ with action $sh$:

$$\begin{aligned} \Pr(H_{i+1}|H_i, sh) &= (1-\rho)(1 - Pr_{at|sh}), \\ \Pr(A_1|H_i, sh) &= (1-\rho)(1-\alpha)^q Pr_{at|sh}, \\ \Pr(P|H_i, sh) &= \rho + (1-\rho)\{1-(1-\alpha)^q\} Pr_{at|sh}. \end{aligned} \quad (6)$$

Moreover, aside from the sensing time, the attacker can still end up in state $P$ during the attack interval. The second part of $\Pr(P|T_i, sh)$ in Eq. 6 represents this situation. The transition probabilities from state $P$ with action $sh$ is,

$$\Pr(H_1|P, sh) = (1-\rho)(1-\Pr_{at|sh}),$$
$$\Pr(A_1|P, sh) = (1-\rho)(1-\alpha)^q \Pr_{at|sh}, \quad (7)$$
$$\Pr(P|P, sh) = \rho + (1-\rho)\{1-(1-\alpha)^q\}\Pr_{at|sh}.$$

In state $A_j$, as the victim device has experienced transmission failures $j$ times in $j$ different channels, it refrains from visiting back to these channels until it successfully finishes the current transmission. Therefore, when an attacker takes action $hop_{freq}$ from state $A_j$, it randomizes its attack sequence, excluding these $j$ channels. Therefore, the probability that the attacker will attack the new operating channel of the victim in the next slot is uniformly distributed over $M-j$ channels. Hence, the probability of an attack is,

$$\Pr_{at|sh,A_j} = \frac{m}{M-j}. \quad (8)$$

The transition probabilities from state $A_j$ with action $sh$ is,

$$\Pr(H_1|A_j, sh) = (1-\rho)(1-P_{at|sh,A_j}),$$
$$\Pr(A_{j+1}|A_j, sh) = (1-\rho)(1-\alpha)^q \Pr_{at|sh,A_j}\{1-\Pr_{det}^j\},$$
$$\Pr(D|A_j, sh) = (1-\rho)(1-\alpha)^q \Pr_{at|sh,A_j} \Pr_{det}^j, \quad (9)$$
$$\Pr(P|A_j, sh) = \rho + (1-\rho)\{1-(1-\alpha)^q\}\Pr_{at|sh,A_j},$$

where $\Pr_{det}^j$ represents the probability of detection by the IDS system. We consider that $\Pr_{det}^j \in R | 0 \leq \Pr_{det}^j \leq 1]$ is a non-decreasing function of $j$. The intuition is that the more an attacker perpetrates, its exposure increases. In this paper, we consider $\Pr_{det}^j = (j-1)/(j-1+c)$, where $c$ is an IDS performance parameter; a lower value of $c$ represents better detection performance by the IDS system.

Now, the transition probabilities with action $mh$ are similar to the action $sh$ only if the frequency assignments at each location are independent. Note that, though action $stay_{loc}+stay_{freq}$ is available, it is a violation of hard-coded attack policy and subject to penalty (i.e., $-F$).

Here, successive attacks increases the probability of a successful attack in the next slot at the current location, i.e.,

$$\Pr(H_1|A_j, sh) > \Pr(H_1|A_{j+1}, sh). \quad (10)$$

## V. PERFORMANCE EVALUATION

To assess the adverse effect of the PACMAN attack, we conduct simulations both in NS3 and MATLAB. Using the simulation in NS3, we depict the performance degradation of a single zone due to the malicious behavior of a Wi-Fi AP. In MATLAB, we implement the proposed MDP to evaluate the attacker's ability to locate critical areas for industrial automation. In the following, we will discuss the implication of the attack based on the analytical and simulation results.

### A. Impact on IAS

Using NS3, we evaluate the impact of the PACMAN attack in a particular zone while presenting the global scenario when the malicious entity is active. To illustrate this, we have considered 3 hexagonal zones in which there would be 3 small BSs associated with 4 UEs. In one of the zones, we have deployed a Wi-Fi AP within the sensing region of the corresponding BS. We consider three scenarios: a non-attack scenario with no active Wi-Fi, a non-attack scenario with active Wi-Fi, and an attack scenario with malicious Wi-Fi. The simulation illustrates half of the run-time (40 seconds) in the benign scenario and the rest of the half in malicious. For the saturated traffic scenario, we have considered a UDP application running at a rate of 155 Mbps with a packet size of 1 KB. For our evaluation, we are considering normalized throughput and delay. Based on Fig. 5(a), a significant increase (146.37% at the end of simulation) in delay can be viewed while comparing to the both in presence and absence of benign Wi-Fi AP. This is because the malicious backoff technique used by the attacker causes a large increase in channel access delay for the victim devices, which has a negative impact on the time-sensitive critical applications of IAS. In terms of throughput as presented in Fig. 5(b), the impact is also sizable (30.56% at the end of simulation) and can potentially impact applications with high throughput demand in IAS.

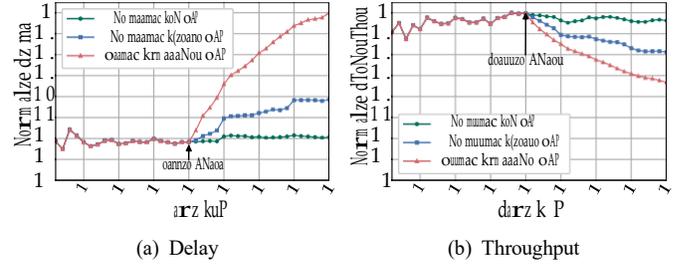

(a) Delay  (b) Throughput

Fig. 5: Impact of the PACMAN attack in a particular zone.

### B. Comparison of Global and Zonal Impact

Using the previously mentioned simulation environment, we have also conducted a comparison of the zonal and global impact of the PACMAN attack. In Fig. 6(a), a significant increase in the victim's zone in terms of average delay is noticed with comparison to the non-attack scenario (the slope is rising rapidly). However, in terms of global impact, the slope is not very steep and the impact is noticeable from 30-35 seconds of the simulation (10-15 seconds after the attack started). In terms of throughput, as depicted in Fig. 6(b), this impact is completely undetectable (5% drop from the non attack scenario) while in the zone the impact is more (18.15% drop from the non attack scenario). If we consider more zones in the attack surface, these impact would be more untraceable from global perspective.

### C. Long Term Impact

From simulation, we can observe a long impact of the attack. In case of delay, as presented in Figures 5(a) and 6(a), the line depicting delay in attack scenario is not converging and taking a long time to reach the maximum delay. The similar phenomenon is also visible in terms of throughput, as of Figures 5(b) and 6(b), where the value did not reach minimum point even after 20 seconds. This mean, it would take a longer time to reach maximum impact of the attack which might be a key to detect such anomaly in the environment.

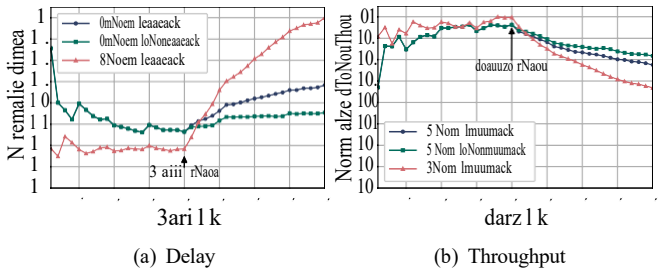

(a) Delay  (b) Throughput

Fig. 6: Comparison of the global and zonal impact of PACMAN attack

## D. Steady-State Sojourn Time and Optimal Policy

The performance of our proposed MDP-based attack strategy is evaluated through its ability to identify critical physical locations. We run the MDP framework in a 7-sectored physical surface (as shown in Fig. 7(a)) with varying degrees of importance toward the industrial automation system. In Fig. 7(a), the color intensity represents the importance of each sector (relative importance is also shown, e.g., S7 is 7 times more important than S4), the name of each sector is provided at the top, and the numerical values represent the normalized total sojourn time of the attacker at each sector. Here, S7 contains the critical devices, and the attacker spends the most time in that location (i.e., 0.99). Also, the attacker moves between S7 and its neighbor, S6, to avoid detection. Fig. 7(b) exhibits the optimal policy in each sector where the arrow and circle represent the $ms$ and $sh$ actions, respectively. The dominant action is shown in a filled line with translucent color, and the secondary action is shown in a dotted line. For instance, the dominant action in S7 is to $sh$, and the secondary action is $ms$ to S6 to avoid detection.

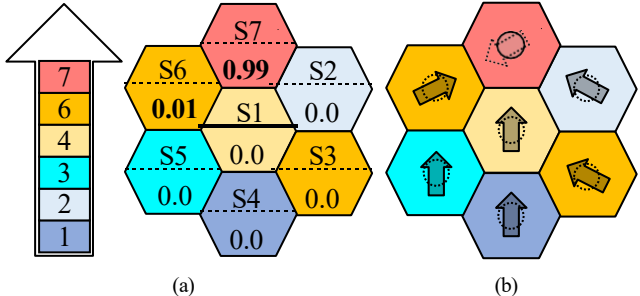

(a)  (b)

Fig. 7: (a) Steady-state sojourn time; (b) Optimal policy.

## VI. CONCLUSION

We proposed a novel mobility-powered Wi-Fi emulation attack model, i.e., PACMAN attack, which exploits the MAC-layer vulnerabilities in a Private 5G-enabled IAS. We also proposed an MDP-based mathematical model to study and assess different dimensions of this attack model. Numerical investigations and simulation results showed that the proposed attack successfully localized the physical locations of critical devices, significantly degraded the performance of the IAS, and compromised network operations. To the best of our knowledge, this is the first work to propose a mobility-powered smart attack against private 5G-enabled IAS.